\documentclass[abstract=on,notitlepage,superscriptaddress,nofootinbib,aps,showpacs,twocolumn,prd]{revtex4}
\usepackage{srcltx}
\usepackage{epsf}
\usepackage{amsfonts}
\usepackage{amsmath}
\usepackage{mathrsfs}
\usepackage{epsfig}
\usepackage{enumerate}
\usepackage{lipsum}
\usepackage{graphicx}
\usepackage{epsf}
\usepackage{subfig}
\usepackage{color}
\usepackage{caption}
\usepackage{subfig}
\usepackage{epstopdf}
\usepackage{float}
\usepackage{dcolumn}
\usepackage{hyperref}
\usepackage{amsthm}

\newcommand{\be}{\begin{equation}}
\newcommand{\ee}{\end{equation}} %\indent}
\newcommand{\eei}{\end{equation}\indent\indent}
\newcommand{\bc}{\begin{center}}
\newcommand{\ec}{\end{center}}
\newcommand{\ber}{\begin{eqnarray*}}
\newcommand{\ear}{\end{eqnarray*}}
\newcommand{\ba}{\begin{array}}
\newcommand{\ea}{\end{array}}

\newcommand{\bea}{\begin{eqnarray}}
\newcommand{\eea}{\end{eqnarray}}
\newcommand{\ei}{\end{itemize}}

\begin{document}

%\preprint{}

\title{What makes a shear-free spherical perfect fluid be inhomogeneous with tidal effects?}



 \author{Jonathan Hakata}\email[]{jonathanhakata@gmail.com}
\affiliation{Astrophysics Research Centre, School of Mathematics, Statistics and Computer Science,	University of KwaZulu--Natal, Private Bag X54001, Durban 4000, South Africa}
\author{Rituparno Goswami}\email[]{Goswami@ukzn.ac.za}
\affiliation{Astrophysics Research Centre, School of Mathematics, Statistics and Computer Science, University of KwaZulu--Natal, Private Bag X54001, Durban 4000, South Africa}
 \author{Chevarra Hansraj}\email[]{chevarrahansraj@gmail.com}
\affiliation{Astrophysics Research Centre, School of Mathematics, Statistics and Computer Science,	University of KwaZulu--Natal, Private Bag X54001, Durban 4000, South Africa}
\author{Sunil D. Maharaj}\email[]{maharaj@ukzn.ac.za}
\affiliation{Astrophysics Research Centre, School of Mathematics, Statistics and Computer Science,	University of KwaZulu--Natal, Private Bag X54001, Durban 4000, South Africa}

%\thanks{}
%\date{\today}

\begin{abstract}
This is an important and natural question as the spacetime shear, inhomogeneity and tidal effects are all intertwined via the Einstein field equations. However, as we show in this paper, such scenarios are possible for limited classes of equations of state that are solutions to a highly non-linear and fourth order differential equation. To show this, we use a covariant semitetrad spacetime decomposition and present a novel geometrical classification of shear-free Locally Rotationally Symmetric (LRS-II) perfect fluid self-gravitating systems, in terms of the covariantly defined fluid acceleration and the fluid expansion. Noteworthily, we deduce the governing differential equation that gives the possible limited equations of state of matter. \\ 

\end{abstract}

%\pacs{04.20.Jb, 04.20.Nr, 04.70.Bw}

\maketitle


\section{Introduction}
 
Spherical shear-free self-gravitating systems of perfect fluids are perhaps the most well studied systems in the historical quest for exact solutions of the Einstein field equations, and as rightly pointed out by \cite{exact}, is rich in rediscoveries. All the known solutions are presented in \cite{McVittie, Kustaan, Wyman, wafo2}, while in all later works these solutions are always contained as special cases. Generally, there are three different approaches to find exact solutions of the field equations for shear-free and spherical perfect fluid distributions. The first approach is via an adhoc ansatz for one of the metric functions, while the second one is to look for symmetries including Lie point symmetries, contact symmetries and Noether symmetries in the field equations. The third approach is more general and rigorous, where the solutions have Painlev\'{e} properties \cite{Wyman}. As a matter of fact, all the known solutions belong to this latter class. 

In this paper, we investigate this well studied system through a more geometrical perspective. In fact, we generalize the spacetime geometry to Locally Rotationally Symmetric (LRS-II) spacetimes \cite{vanElstEllis}, of which spherical symmetry is a subclass. By covariantly decomposing the spacetime using the fluid flow congruence and a preferred spherical congruence which is guaranteed by LRS-II symmetry, we classify all the possible solutions of the field equations in terms of the covariantly defined fluid acceleration and the fluid expansion. 

Shear-free matter distributions have been used to model both static and radiating stars, justifying its assumption. Spherical symmetry simplifies the relativistic equations even further. Spherically symmetric shear-free solutions have been used to model many physical applications leading to many classes of solutions \cite{stephani, srivastava, deng, sussman1, sussman2, maharaj1, havas, maharaj2} where it is demonstrated that zero shear restricts the possible choices of equations of state \cite{sussman1,  sussman2}. Recent applications of this model relate to gravitational collapse \cite{sharif, pinheiro, naidu, charan}; symmetry analysis \cite{wafo, gumede} and methods \cite{msomi, moopanar}; and particular solutions \cite{wagh, bhatti, shah}.

Interestingly, our analysis sheds new light on the class of solutions where fluid acceleration and expansion are strictly non-vanishing. It can easily be seen that the Weyl curvature is also non-vanishing for this class. This class is interesting as it is dynamical and necessarily inhomogeneous but still shear-free. It is well known that spacetime shear and the electric part of the Weyl form a feedback loop with each other. The electric part of the Weyl is a source term for the shear evolution equation while the shear is the source term for the evolution of the electric part of the Weyl. 

\emph{Therefore, constraining one of the positive feedback loop quantities to vanish identically, while the other is strictly non-vanishing, would imply a stringent constraint on other geometrical and thermodynamical quantities in the spacetime.} 

From our analysis, we transparently show that indeed such a stringent constraint exists on the possible equations of state that give rise to this class. We find the governing ODE, which is highly nonlinear, the solution to which gives the possible classes of equations of state that may give rise to this class of solutions. We present a numerical solution for this ODE and clearly show that the usual equations of state that are linear, or a finite combination of power law solutions, will not generate the given class. 

The paper is organized as follows: In the next two sections we discuss the covariant semitetrad decomposition of the spacetime manifold, and the field equations written in terms of geometrical and thermodynamical variables that emerge due to the decomposition. In the subsequent two sections we discuss the special case of shear-free spacetimes, and show how all possible solutions can be classified in terms of the fluid acceleration and the fluid expansion. The next section is dedicated to the interesting class of inhomogeneous and shear-free spacetimes, and the differential equation that governs the equation of state of the perfect fluid. Finally, we conclude our results in the last section.

\section{1+1+2 formalism} \label{formalism}

The 1+1+2 covariant approach, first introduced by Greenberg in \cite{greenberg}, further reviewed by van Elst and Ellis  \cite{vanElstEllis}  and subsequently expanded by Clarkson and Barrett  \cite{clarksonbarrett, clarkson2007} is an extension of the 1+3 formalism \cite{ehlers, ellis1+3}. The latter is one of the most widely used tetrad approaches which formulates the equations of general relativity, using a timelike congruence, as first order differential equations unlike the coordinate approach involving second order partial derivatives of the metric functions. 

In the 1+3 covariant approach the essential `time' coordinate is separated from the 3-`space' so that the metric tensor takes the form
\begin{equation}
g_{ab}=h_{ab}-u_au_b. \label{1+3}
\end{equation}
Two important derivatives are defined. The covariant time derivative along the observers' worldlines, denoted by `\,${^{\cdot}}$\,', is defined using the vector ${u^{a}}$, as
\begin{equation} 
\dot{Z}^{a ... b}{}_{c ... d} = u^{e}\nabla_{e} Z^{a ... b}{}_{c ... d},
\end{equation} 
for any tensor ${Z^{a...b}{}_{c...d}}$. The fully orthogonally projected covariant spatial derivative, denoted by `\,${D}$\,', is defined using the spatial projection tensor ${h_{ab}}$, as
\begin{equation}
D_{e} Z^{a...b}{}_{c...d} = h^r{}_{e} h^p{}_{c}... h^q{}_{d} h^a{}_{f}... h^b{}_{g}\nabla_{r} Z^{f...g}{}_{p...q},
\end{equation}
with total projection on all the free indices. 
In the context of Locally Rotationally Symmetric class II (LRS-II) spacetimes \cite{vanElstEllis}, the geometrical quantities defined for the timelike congruence are the expansion scalar $(\Theta = D_{a} u^{a} )$, the acceleration 3-vector $(\dot{u}^a = u^{b}\nabla_{b}u^a )$ and the shear 3-tensor $[\sigma_{ab} =  \left(h^{c}{}_{(a} h^{d}{}_{b)} - \frac{1}{3}h_{ab} h^{cd}\right)D_{c} u_{d}]$. The timelike congruence uniquely defines the electric part of the Weyl tensor $(E_{ab} = C_{abcd} u^{c} u^{d} = E_{<ab>})$ and the magnetic part vanishes identically. The energy momentum tensor is also decomposed according to the timelike congruence to produce the energy density $(\mu = T_{ab} u^{a} u^{b})$, the isotropic pressure $(p = \frac{1}{3} h_{ab} T^{ab})$, the heat flux 3-vector $(q_a = q_{<a>} = - h^{c}{}_{a} T_{cd} u^{d})$ and the anisotropic stress 3-tensor $(\pi_{ab} = T_{cd} h^c{}_{<a} h^d{}_{b>})$. Angle brackets denote orthogonal projections of covariant time derivatives along ${u^{a}}$ as well as represent the projected, symmetric and trace-free part of tensors as follows
\begin{eqnarray}
v_{<a>} &=& h^{b}{}_{a}\dot{v}_{b},  \nonumber\\
Z_{<ab>} &=& \left(h^{c}{}_{(a} h^{d}{}_{b)} - \frac{1}{3}h_{ab} h^{cd}\right) Z_{cd}.
\end{eqnarray} 
All these quantities have a direct geometrical and physical meaning and are described by tensorial quantities that remain valid in all coordinate systems which is the principal advantage of using spacetime decomposition. 

A further decomposition of the 1+3 quantities gives rise to the useful 1+1+2 set of variables featured in  \cite{clarksonbarrett}. The splitting of the LRS-II spacetime is performed with respect to the timelike unit vector $u^a$ ($u^au_a = -1$) as well as the isolation of a preferred spatial direction $e^a (e^ae_a = 1)$ chosen orthogonal to $u^a (u^a e_a = 0)$. The metric tensor \eqref{1+3} is decomposed further into
\begin{equation}
g_{ab}=-u_a u_b+e_a e_b+N_{ab},
\end{equation}
where $N_{ab}$ is the 2-dimensional metric on the spherical 2-shell. We introduce two new derivatives for any tensor ${\Psi_{a...b}{}^{c...d}}$:
\begin{eqnarray}
\label{hatderiv}
\hat{\Psi}_{a...b}{}^{c...d} &\equiv& e^{f} D_{f} \Psi_{a...b}{}^{c...d}, \\
\label{deltaderiv}
\delta_{f}\Psi_{a...b}{}^{c...d} &\equiv& N_{f}{}^{j} N_{a}{}^{l} ... N_{b}{}^{g} N_{h}{}^{c} ... N_{i}{}^{d}  D_{j}\Psi_{l...g}{}^{h...i}, 
\end{eqnarray}
defined by the spatial congruence ${e^{a}}$. The hat-derivative (\ref{hatderiv}) is the spatial derivative along the ${e^{a}}$ vector field in the surfaces orthogonal to ${u^{a}}$, and the delta-derivative (\ref{deltaderiv}) is the projected spatial derivative onto the 2-sheet, with projection on every free index. In a similar way, there are geometrical quantities generated for the preferred spatial congruence with the only non-vanishing quantity for LRS-II spacetimes being the 2-sheet volume expansion $(\phi = \delta_a e^a)$. Using the preferred spatial congruence we can then extract a set of covariant scalars that completely govern the dynamics of the system in the following manner
\begin{eqnarray}
\mathcal{A} &=& \dot{u}^a e_a, \quad \Sigma = \sigma_{ab}e^a e^b, \mathcal{E} = E_{ab}e^a e^b, \nonumber\\
 Q &=& q^a e_a, \quad \Pi = \pi_{ab}e^a e^b.
\end{eqnarray}

Evidently the 1+1+2 decomposition method is well suited to spacetimes that have a preferred spatial direction such as LRS-II spacetimes where $u^a$ and $e^a$ are hypersurface orthogonal. These spacetimes have the inherent property that there exists a unique preferred spacial direction at each point that creates a local axis of symmetry. Hence all the physics and geometry of the spacetime are described by well defined kinematic and dynamic scalar variables that generate the field equations. The 1+1+2 covariant method has generated new results in studies involving LRS spacetimes \cite{sayuri}, the Kerr spacetime \cite{kerr}, the Vaidya spacetime \cite{vaidya} and a general spacetime admitting conformal symmetry \cite{CKV}.

%%%%%%%%%%%%%%%%%%%%%%%%%%%%%%%%%%%%%%%%%%%%%%%%%%%%%%%%%%%%%%%%%%%%%%%%%%%%%%%%%%%%%%%%%%%%%%%%%%%%%%%%%%%%%%%%%%%%%%
%%%%%%%%%%%%%%%%%%%%%%%%%%%%%%%%%%%%%%%%%%%%%%%%%%%%%%%%%%%%%%%%%%%%%%%%%%%%%%%%%%%%%%%%%%%%%%%%%%%%%%%%%%%%%%%%%%%%%%

\section{Field equations for LRS-II spacetimes}

As pointed out by Goswami and Ellis \cite{birkhoff}, the only variables in the 1+1+2 formalism which characterize the kinematics are  
\begin{equation} \label{setD}
\mathcal{D} = \left\{\mathcal{A},\Theta,\phi,\Sigma,\mathcal{E},\mu,p,\Pi,Q\right\}.
\end{equation}
For these variables the propagation, evolution and constraint equations can be derived as
\begin{eqnarray}
\label{LRSstart}
	\hat{\phi} &=& -\frac{1}{2}\phi^2+\left(\frac{1}{3}\Theta+\Sigma\right)\left(\frac{2}{3}\Theta - \Sigma\right) \nonumber\\
	& & -\frac{2}{3}\mu-\frac{1}{2}\Pi-\mathcal{E},\label{A}\\	
	\hat{\Sigma}-\frac{2}{3}\hat{\Theta} &=&-\frac{2}{3}\phi\Sigma-Q,\label{B}\\	
	\hat{\mathcal{E}}-\frac{1}{3}\hat{\mu}+\frac{1}{2}\hat{\Pi} &=&-\frac{3}{2}\phi\left(\mathcal{E} +\frac{1}{2}\Pi\right) 
	\nonumber\\ && +\left(\frac{1}{2}\Sigma-\frac{1}{3}\Theta\right)Q,\label{C}\\	
	\dot{\phi}&=&-\left(\Sigma-\frac{2}{3}\Theta\right)\left(\mathcal{A}-\frac{1}{2}\phi\right)+Q,\label{D}\\	
	\dot{\Sigma}-\frac{2}{3}\dot{\Theta}&=&-\mathcal{A}\phi+2\left(\frac{1}{3}\Theta-\frac{1}{2}\Sigma\right)^2 -\mathcal{E}
	\nonumber\\ & & +\frac{1}{3}(\mu+3p)+\frac{1}{2}\Pi,\label{E} \\	
	\dot{\mathcal{E}}-\frac{1}{3}\dot{\mu}+\frac{1}{2}\dot{\Pi}&=&\left(\frac{3}{2}\Sigma-\Theta\right)\mathcal{E}+\frac{1}{4}\left(\Sigma-\frac{2}{3}\Theta\right)\Pi
	\nonumber\\ && +\frac{1}{2}\phi Q-\frac{1}{2}(\mu+p)\left(\Sigma-\frac{2}{3}\Theta\right),\label{F}\\	
	\dot{\mu}+\hat{Q}&=& -\Theta(\mu+p) -\frac{3}{2}\Sigma\Pi \nonumber\\
	 && -(\phi+2\mathcal{A})Q,\label{G}\\	
	\dot{Q}+\hat{p}+\hat{\Pi}&=&-\left(\frac{3}{2}\phi+\mathcal{A}\right)\Pi -(\mu+p)\mathcal{A} \nonumber\\
	&& -\left(\frac{4}{3}\Theta+\Sigma\right)Q,\label{H}\\	
	\label{LRSend}
	\hat{\mathcal{A}}-\dot{\Theta}&=&-(\mathcal{A}+\phi)\mathcal{A}+\frac{1}{3}\Theta^2 \nonumber\\
	& &+\frac{3}{2}\Sigma^2+\frac{1}{2}\left(\mu+3p\right)\label{I},
\end{eqnarray}
according to \cite{clarkson2007}. The field equations are not a closed set of equations because there are no explicit equations for $\dot{\mathcal{A}}$ (evolution equation for $\mathcal{A}$) and $\hat{\Theta}$ (propagation equation for $\Theta$). Hence  an equation of state, governing the thermodynamical quantities, is needed of the general form $F(\mu,p,\Pi,Q)=0$. This form becomes simplified for a shear-free perfect fluid model where there are no shear stresses or heat flux $(\Pi = Q =0$). Hence our equation of state takes the form
 \begin{equation}
 p=p(\mu), \label{eos}
 \end{equation}
 where the isotropic pressure $p$ is a function of $\mu$, the effective energy density. At this stage, we need to point out that the derivative operators `$\delta^a$', `${^\cdot{}}$' and ` $\hat{}$ ' do not commute and give rise to an interesting result later. Instead, according to \cite{clarkson2007}, the commutation relation when acting on some scalar $\beta$ for our imposed LRS-II spacetime conditions is given by 
\begin{align}
     \hat{\dot{\beta}}-\dot{\hat{\beta}}=-\mathcal{A}\dot{\beta}+
     \left(\frac{1}{3}\Theta+\Sigma\right)\hat{\beta},\label{comm_reln}
\end{align}
which will be used in our calculations in the subsequent sections.

%%%%%%%%%%%%%%%%%%%%%%%%%%%%%%%%%%%%%%%%%%%%%%%%%%%%%%%%%%%%%%%%%%%%%%%%%%%%%%%%%%%%%%%%%%%%%%%%%%%%%%%%%%%%%%%%%%%%%%
%%%%%%%%%%%%%%%%%%%%%%%%%%%%%%%%%%%%%%%%%%%%%%%%%%%%%%%%%%%%%%%%%%%%%%%%%%%%%%%%%%%%%%%%%%%%%%%%%%%%%%%%%%%%%%%%%%%%%%

\section{Shear-free perfect fluid LRS-II  equations}
Thus far and later in this paper, we consider a shear-free  perfect fluid LRS-II  spacetime with an equation of state $\eqref{eos}$. Perfect fluids were studied in the context of the 1+3 decomposition method \cite{vanElstEllis} and for the 1+1+2 decomposition method \cite{clarkson2007}. Such a fluid appears to be a good description of the observed universe on a large scale. The absence of  shear stresses, viscosity and heat conduction is a great advantage as the relativistic equations become simpler. For this shear-free  perfect fluid LRS-II model we have
\begin{equation}
\Sigma=Q=\Pi=0. \label{conditions}
\end{equation} 
Hence the above general LRS-II equations \eqref{LRSstart}-\eqref{LRSend} simplify to
\begin{eqnarray}
	\hat{\phi} &=& -\frac{1}{2}\phi^2+\frac{2}{9}\Theta^2-\frac{2}{3}\mu-\mathcal{E},\label{A1}\\	
	\hat{\Theta} &=& 0,\label{B1}\\	
	\hat{\mathcal{E}}-\frac{1}{3}\hat{\mu}&=&-\frac{3}{2}\phi\mathcal{E},\label{C1}\\	
	\dot{\phi}&=&\frac{2}{3}\Theta\mathcal{A}-\frac{1}{3}\Theta\phi,\label{D1}\\	
	-\frac{2}{3}\dot{\Theta}&=&-\mathcal{A}\phi+\frac{2}{9}\Theta^2+\frac{1}{3}(\mu+3p)-\mathcal{E},\label{E1} \\	
	\dot{\mathcal{E}}-\frac{1}{3}\dot{\mu}&=&-\Theta\mathcal{E}+\frac{1}{3}\Theta(\mu+p),\label{F1}\\	
	\dot{\mu} &=& -\Theta(\mu+p),\label{G1}\\	
	\hat{p} &=&-\mathcal{A}(\mu+p),\label{H1}\\	
	\hat{\mathcal{A}}-\dot{\Theta}&=&-(\mathcal{A}+\phi)\mathcal{A}+\frac{1}{3}\Theta^2+\frac{1}{2}(\mu+3p)\label{I1}.
\end{eqnarray}

We immediately note that due to our imposed conditions \eqref{conditions} we obtain an explicit expression for $\hat{\Theta}$ in \eqref{B1}; however, we still do not have an explicit expression for $\dot{\mathcal{A}}$. At this point, we make $\mathcal{A}$ the central focus of our study and use it to classify shear-free perfect fluid LRS-II  solutions in the next section.

%%%%%%%%%%%%%%%%%%%%%%%%%%%%%%%%%%%%%%%%%%%%%%%%%%%%%%%%%%%%%%%%%%%%%%%%%%%%%%%%%%%%%%%%%%%%%%%%%%%%%%%%%%%%%%%%%%%%%%
%%%%%%%%%%%%%%%%%%%%%%%%%%%%%%%%%%%%%%%%%%%%%%%%%%%%%%%%%%%%%%%%%%%%%%%%%%%%%%%%%%%%%%%%%%%%%%%%%%%%%%%%%%%%%%%%%%%%%%

\section{Classes of shear-free perfect fluid LRS-II  solutions}
For the equations outlined in the preceding section, we obtain various classes of shear-free perfect fluid LRS-II solutions characterized by the acceleration scalar $\mathcal{A}$. First, it is pertinent to consider the simplest case where acceleration is absent.

%%%%%%%%%%%%%%%%%%%%%%%%%%%%%%%%%%%%%%%%%%%%%%%%%%%%%%%%%%%%%%%%%%%%%%%%%%%%%%%%%%%%%%%%%%%%%%%%%%%%%%%%%%%%%%%%%%%%%%

\subsection{LRS-II spacetime with no acceleration: $\mathcal{A} = 0$}\label{no_accn}

In this case, we immediately see from equation (\ref{H1}) that the pressure must be homogeneous with $\hat{p}=0$. Now comparing equations (\ref{E1}) and (\ref{I1}), implies $  \mathcal{E}=0$.
Equation (\ref{C1}) then implies $\hat{\mu}=0$, and hence this scenario describes a dynamic and homogeneous matter distribution.
Since we have $\mathcal{A} = \mathcal{E} = 0$ and the matter is homogeneous, the spacetime has to strictly be Friedmann-Lema\^itre-Robertson-Walker (FLRW). It is also important to note that from the general LRS-II field equations \eqref{A}-\eqref{I}, we cannot have a perfect fluid solution where $\Theta = \mathcal{A}=0$, so we do not consider that solution branch.

%%%%%%%%%%%%%%%%%%%%%%%%%%%%%%%%%%%%%%%%%%%%%%%%%%%%%%%%%%%%%%%%%%%%%%%%%%%%%%%%%%%%%%%%%%%%%%%%%%%%%%%%%%%%%%%%%%%%%%

\subsection{LRS-II spacetime with acceleration: $\mathcal{A} \neq 0$}
From the equations \eqref{A1}-\eqref{I1}, we noted earlier that we do not have an explicit equation for  $\dot{\mathcal{A}}$. However with the imposed equation of state \eqref{eos}, we can obtain this crucial evolution equation so that we have a closed set of equations, and we now describe this process in detail. 

Considering the definition of the isentropic speed of sound $c_s^2 = \left(\partial p/ \partial \mu\right)_{s = \text{constant}}$, we note that we can write
\begin{eqnarray}
\hat{p} &=& c_s^2 \hat{\mu} = \frac{\partial p}{\partial \mu}\,\hat{\mu},\\ \label{phat}
 \dot{p} &=& c_s^2\dot{\mu} =\frac{\partial p}{\partial \mu}\,\dot{\mu} . \label{pdot}
\end{eqnarray}
Therefore, from equation \eqref{H1},
\begin{eqnarray}
	  \mathcal{A} &=& -\frac{\hat{p}}{(\mu+p)},\nonumber \\
	\implies \dot{\mathcal{A}} &=& -\frac{\dot{\hat{p}}}{(\mu+p)}+\mathcal{A}\Theta\left(1+\frac{\partial p}{\partial \mu}\right), \label{K1}
\end{eqnarray}
using  \eqref{G1}  and \eqref{pdot} for simplification. Obtaining an expression for $\dot{\hat{p}}$ from the commutation relation \eqref{comm_reln} and substituting into equation \eqref{K1} we get
\begin{equation}
	\dot{\mathcal{A}}=\mathcal{A}\Theta\left(\frac{\partial p}{\partial \mu}-\frac{1}{3}-\frac{(\mu+p)}{c_s^2}\frac{\partial^2 p}{\partial \mu^2}\right)=\mathcal{A}\Theta\mathcal{F},\label{K}
\end{equation}
after simplification using \eqref{phat} and where we have set
\begin{equation}
\mathcal{F} \equiv \mathcal{F}(\mu) = \left(\frac{\partial p}{\partial \mu}-\frac{1}{3}-\frac{(\mu+p)}{c_s^2}\frac{\partial^2 p}{\partial \mu^2}\right).
\end{equation}

We now investigate how the obtained acceleration evolution equation \eqref{K} is compatible with the system from the integrability condition; to achieve this we make use of the commutation relation \eqref{comm_reln}. According to the left hand side of equation \eqref{comm_reln}, we need to obtain $\hat{\dot{\mathcal{A}}}$ and $\dot{\hat{\mathcal{A}}}$. Using \eqref{K}, we have
\begin{eqnarray}
	 \hat{\dot{\mathcal{A}}} &=&\hat{\mathcal{A}}\Theta\mathcal{F}+\mathcal{A}\hat{\Theta}\mathcal{F}+\mathcal{A}\Theta\hat{\mathcal{F}}\nonumber \\
	&=&\Theta\mathcal{F}\left(-\mathcal{A}^2+\frac{1}{2}\mathcal{A}\phi+\frac{3}{2}\mathcal{E}\right)
	\nonumber\\ && -\mathcal{A}^2\Theta\frac{(\mu+p)}{c_s^2}\mathcal{F}', \label{L}
\end{eqnarray}
where $\mathcal{F}'=\frac{\partial \mathcal{F}(\mu)}{\partial \mu}$.
Using \eqref{E1} in the Raychaudhuri equation \eqref{I1} we get
\begin{eqnarray}
	\hat{\mathcal{A}}&=&-\mathcal{A}^2+\frac{1}{2}\mathcal{A}\phi+\frac{3}{2}\mathcal{E}, \nonumber \\
	\implies \dot{\hat{\mathcal{A}}} &=& -2\mathcal{A}^2\Theta\mathcal{F}+\frac{1}{3}\mathcal{A}^2\Theta-\frac{1}{6}\mathcal{A}\Theta\phi \nonumber\\ &&
	 +\frac{1}{2}\mathcal{A}\Theta\phi\mathcal{F}-\frac{3}{2}\Theta\mathcal{E}. \label{M} 
\end{eqnarray}
Then subtracting \eqref{M} from \eqref{L}, as required by the commutation relation \eqref{comm_reln}, yields
\begin{eqnarray}
	\hat{\dot{\mathcal{A}}}-\dot{\hat{\mathcal{A}}} &=&  \mathcal{A}^2\Theta\mathcal{F}+\frac{3}{2}\Theta\mathcal{E}\mathcal{F}  -\mathcal{A}^2\Theta\frac{(\mu+p)}{c_s^2}\mathcal{F}' \nonumber\\
	&&-\frac{1}{3}\mathcal{A}^2\Theta+\frac{1}{6}\mathcal{A}\Theta\phi+\frac{3}{2}\Theta\mathcal{E}.\label{N}
\end{eqnarray}
However according to the commutation relation \eqref{comm_reln}, we expected to obtain
\begin{eqnarray}
\hat{\dot{\mathcal{A}}}-\dot{\hat{\mathcal{A}}} &=& -\mathcal{A}\dot{\mathcal{A}}+\frac{1}{3}\Theta\hat{\mathcal{A}} \nonumber\\
	&=&-\mathcal{A}^2\Theta\mathcal{F}-\frac{1}{3}\mathcal{A}^2\Theta
	+\frac{1}{6}\mathcal{A}\Theta\phi +\frac{1}{2}\Theta\mathcal{E}.\label{O}
\end{eqnarray}
We immediately notice the interesting discrepancy between \eqref{N} and \eqref{O}. They do not match and this needs to be resolved for the existence of solutions. The condition for $\dot{\mathcal{A}}$ to be compatible with the system is obtained by setting the difference between \eqref{N} and \eqref{O} to zero as follows
\begin{equation}
2\mathcal{A}^2\Theta\mathcal{F}+\frac{3}{2}\Theta\mathcal{E}\mathcal{F}-\mathcal{A}^2\Theta\frac{(\mu+p)}{c_s^2}\mathcal{F}'+\Theta\mathcal{E} = 0,
\end{equation}
which simplifies to
\begin{eqnarray}	
 \Theta\left[\mathcal{A}^2\left(2\mathcal{F}-\frac{(\mu+p)}{c_s^2}\mathcal{F}'\right)  +\mathcal{E}\left(\frac{3}{2}\mathcal{F}+1\right)\right] = 0. \label{P}
\end{eqnarray}
Equation \eqref{P} implies that either $\Theta = 0$ or the square-bracketed term is zero.

%%%%%%%%%%%%%%%%%%%%%%%%%%%%%%%%%%%%%%%%%%%%%%%%%%%%%%%%%%%

\subsubsection{Case 1: Static fluid $\left(\Theta = 0\right)$}
Since we have $\Sigma = Q =\Pi=\Theta = 0$ then the spacetime is necessarily static. This means that any reasonable equation of state of the form $p=p(\mu)$ will solve the system.

%%%%%%%%%%%%%%%%%%%%%%%%%%%%%%%%%%%%%%%%%%%%%%%%%%%%%%%%%%%

\subsubsection{Case 2: Nonstatic fluid $\left(\Theta \neq 0\right)$}
\begin{widetext}

 \begin{figure}
\begin{centering}

\includegraphics[width=0.80\textwidth]{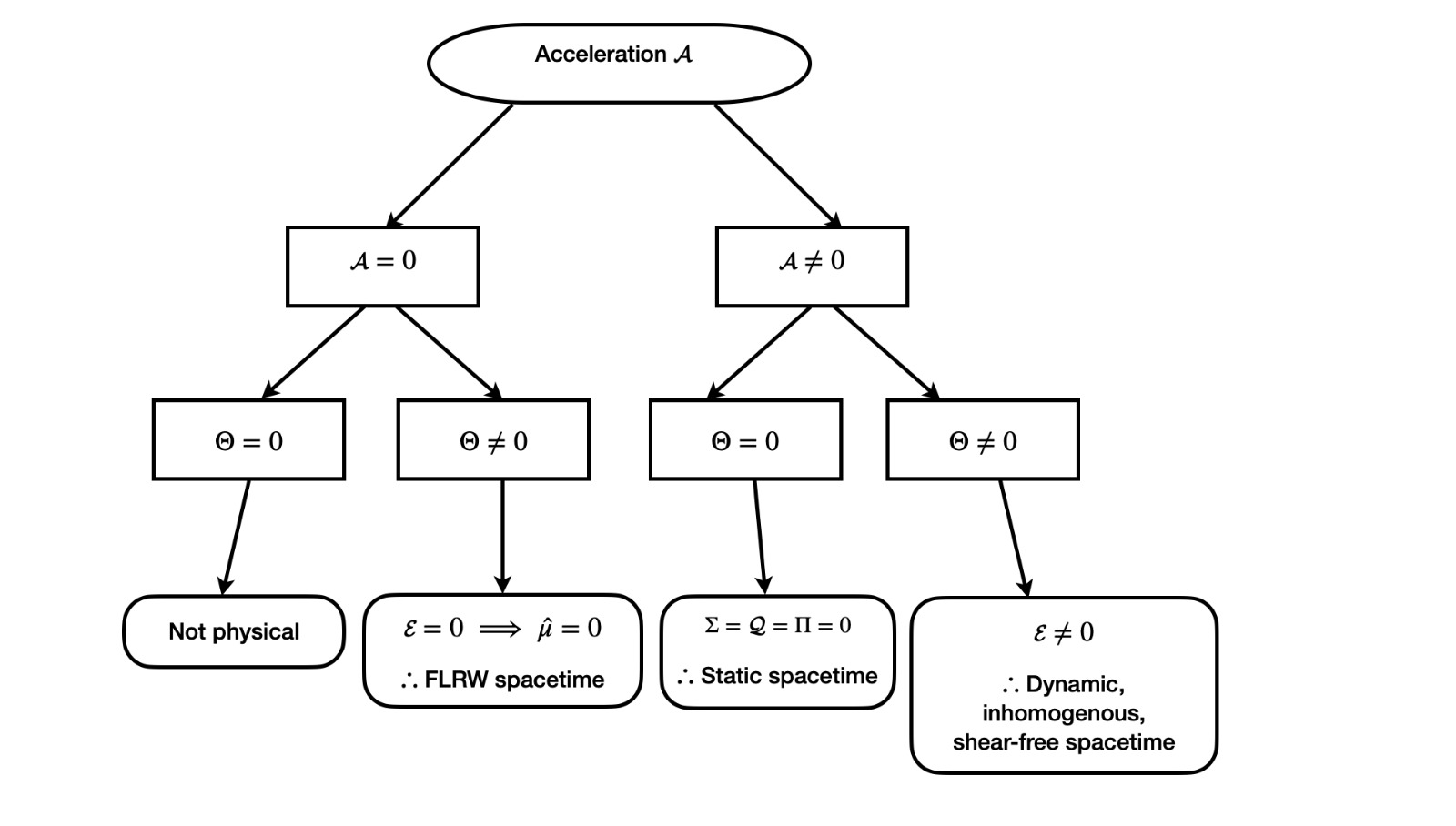}

\caption{Diagram illustrating the various configurations of the property of acceleration and the spacetime classification.}
\label{fig:1}
% Give a unique label
\end{centering}
\end{figure}

\end{widetext}

In this case $\Theta$ is strictly not equal to zero. Hence the square-bracketed term in \eqref{P} has to be zero, that is
\begin{equation}
    \left[\mathcal{A}^2\left(2\mathcal{F}-\frac{(\mu+p)}{c_s^2}\mathcal{F}'\right)+\mathcal{E}\left(\frac{3}{2}\mathcal{F}+1\right)\right]=0.
\end{equation}
Two possible subcases arise.
\begin{itemize}
	\item \textbf{Subcase 1:} 
	If both $\mathcal{A}=\mathcal{E}=0$ then the spacetime is necessarily Friedmann-Lema\^itre-Robertson-Walker (FLRW). This type reduces to the case contained in section \ref{no_accn}.
	\item \textbf{Subcase 2:} If $\mathcal{A}$ and $\mathcal{E}$ are well defined and both nonzero then from \eqref{P} we have
	 \begin{eqnarray}
	 	&&\mathcal{A}^2\left(2\mathcal{F}-\frac{(\mu+p)}{c_s^2}\mathcal{F}'\right)+\mathcal{E}\left(\frac{3}{2}\mathcal{F}+1\right)= 0, \nonumber \\
	 &&	\implies \frac{\mathcal{E}}{\mathcal{A}^2}= \frac{\frac{(\mu+p)}{c_s^2}\mathcal{F}'-2\mathcal{F}}{\frac{3}{2}\mathcal{F}+1}. \label{Q}
	 \end{eqnarray}
 \end{itemize}
Subcase 2 gives us a new constraint equation \eqref{Q} that needs to be consistently satisfied by the equation of state. Now for this constraint equation to time-evolve consistently, its time derivative must also be zero, and evolving \eqref{P} once yields
\begin{eqnarray} 
 &&\Theta\left[\mathcal{A}^2\left(2\mathcal{F}\mathcal{G}-(\mu+p)\mathcal{G}'\right) \right.
\nonumber\\&& 
\left.  - \mathcal{E}\left(\frac{3}{2}\mathcal{F}+1+\frac{3}{2}(\mu+p)\mathcal{F}'\right)\right]=0, \label{R}
\end{eqnarray}
where we have set
\begin{equation} 
	\mathcal{G} \equiv \mathcal{G}(\mu) = 2\mathcal{F}-\frac{(\mu+p)}{c_s^2}\mathcal{F}',
\end{equation}
and  $\mathcal{G}'=\frac{\partial \mathcal{G}(\mu)}{\partial \mu}$. As we shall see in the next section, any further time evolution of the above constraint is not required as the constraints (\ref{Q}) and (\ref{R}) give us the governing differential equation for the equation of state of the matter that gives rise to this given subcase. We summarize the findings in this section in a diagram depicting the various classes of spherically symmetric shear-free perfect fluid solutions characterized by the acceleration scalar $\mathcal{A}$ in Figure \ref{fig:1}.
 
%%%%%%%%%%%%%%%%%%%%%%%%%%%%%%%%%%%%%%%%%%%%%%%%%%%%%%%%%%%%%%%%%%%%%%%%%%%%%%%%%%%%%%%%%%%%%%%%%%%%%%%%%%%%%%%%%%%%%%
%%%%%%%%%%%%%%%%%%%%%%%%%%%%%%%%%%%%%%%%%%%%%%%%%%%%%%%%%%%%%%%%%%%%%%%%%%%%%%%%%%%%%%%%%%%%%%%%%%%%%%%%%%%%%%%%%%%%%%

\section{Dynamic and inhomogeneous fluid}
In the event that $\Theta \neq 0$ and $\mathcal{A} \neq 0$,  setting the square-bracketed term in equation \eqref{R} to zero yields
\begin{equation}
	\frac{\mathcal{E}}{\mathcal{A}^2}=\frac{2\mathcal{F}\mathcal{G}-(\mu+p)\mathcal{G}'}{\frac{3}{2}\mathcal{F}+1+\frac{3}{2}(\mu+p)\mathcal{F}'}. \label{S}
\end{equation}
 For consistency, equations \eqref{Q} and \eqref{S} need to be equivalent and this condition produces a master differential equation for the equation of state given by
\begin{eqnarray}
	\frac{\frac{(\mu+p)}{c_s^2}\mathcal{F}'-2\mathcal{F}}{\frac{3}{2}\mathcal{F}+1} = \frac{2\mathcal{F}\mathcal{G}-(\mu+p)\mathcal{G}'}{\frac{3}{2}\mathcal{F}+1+\frac{3}{2}(\mu+p)\mathcal{F}'}, \nonumber
\end{eqnarray}
which simplifies to
\begin{widetext}
\begin{eqnarray}\label{master}
	 \left(\frac{(\mu+p)}{c_s^2}\mathcal{F}'-2\mathcal{F}\right)\left(\frac{3}{2}\mathcal{F}+\frac{3}{2}(\mu+p)\mathcal{F}'  + 1 \right) -\left(\frac{3}{2}\mathcal{F}+1\right)\left(2\mathcal{F}\mathcal{G}-(\mu+p)\mathcal{G}'\right)  = 0.
\end{eqnarray}
\end{widetext}
Equation \eqref{master} imposes a constraint on the equation of state. Finding an equation of state which is a solution to \eqref{master} will then necessarily make the two constraints (\ref{Q}) and (\ref{R}) identically compatible. Since these two constraints are identically satisfied by the equation of state (which is the solution of the above equation), any further time evolution of these constraints will also be identically satisfied and the system of equations is well posed. From equation \eqref{S}, both $\Theta$ and $\mathcal{A}$ are nonzero hence $\hat{\mu}$ and $\hat{p}$ are also nonzero and so these solutions are necessarily dynamic and inhomogeneous but shear-free.

The resulting differential equation \eqref{master} is not autonomous as it stands. However, it can be transformed into that type via a change of variables. Letting $X=(\mu+p)$ yields the autonomous ordinary differential equation given by

\begin{widetext}
\begin{eqnarray}
\label{autoODE}
&&-\frac{3}{2} \left(X'-1\right) \left(XX'' -\left(X'\right)^{2}+\frac{5}{3}X'-\frac{2}{3}\right) X^{3} X''''   + \frac{3}{2}XX'''\left\{-\left(X -2 X' +2\right) \left(XX''\right)^{2}  \right.  \nonumber\\
&& +\left(X'-1\right) \left[-4 \left(X'\right)^{2} +\left(4 X +\frac{16}{3}\right)X' +X^{2}-\frac{11}{3}X-\frac{4}{3}\right] XX'' \nonumber\\
&&  \left. -4\left(X'-1\right)^{2} \left(X'-\frac{2}{3}\right) \left(-\frac{1}{2}\left(X'\right)^{2}+\left(X+\frac{1}{2}\right)X' -\frac{7}{6}X\right) \right\} -\frac{3}{2} \left(-13 X' +X+9\right) \left(XX''\right)^{3}\nonumber\\
&&  -6\frac{\left(XX''\right)^{4}}{X'-1}  -3 \left(X'-1\right) \left(8 \left(X'\right)^{2}+\left(X -12\right)X' -\frac{3}{2} X+\frac{35}{9}\right) \left(XX''\right)^{2} \nonumber\\
&& +3 \left(X'-1\right)^{2} \left(6 \left(X'\right)^{3}+\left(X-\frac{95}{6}\right) \left(X'\right)^{2}+\left(-\frac{7}{3}X + \frac{233}{18}\right)X' +\frac{4}{3} X - \frac{29}{9}\right) XX'' \nonumber\\
&& -6 \left(X'-1\right)^{4} \left(X'-\frac{2}{3}\right) \left(X'-\frac{4}{3}\right) \left(X'-\frac{5}{6}\right)=0.
\end{eqnarray}
\end{widetext}

This equation is not satisfied by either linear or power laws of the form $p=\kappa\mu$ or $p=\kappa\mu^\gamma,(\gamma>2)$ respectively. This infers that it has a complicated solution and so carrying out a numerical analysis on the mathematical software Maple \cite{maple} yields the solution given in Figure 2.
\begin{figure}[ht!]
	\centering	\includegraphics[width=0.3\textwidth]{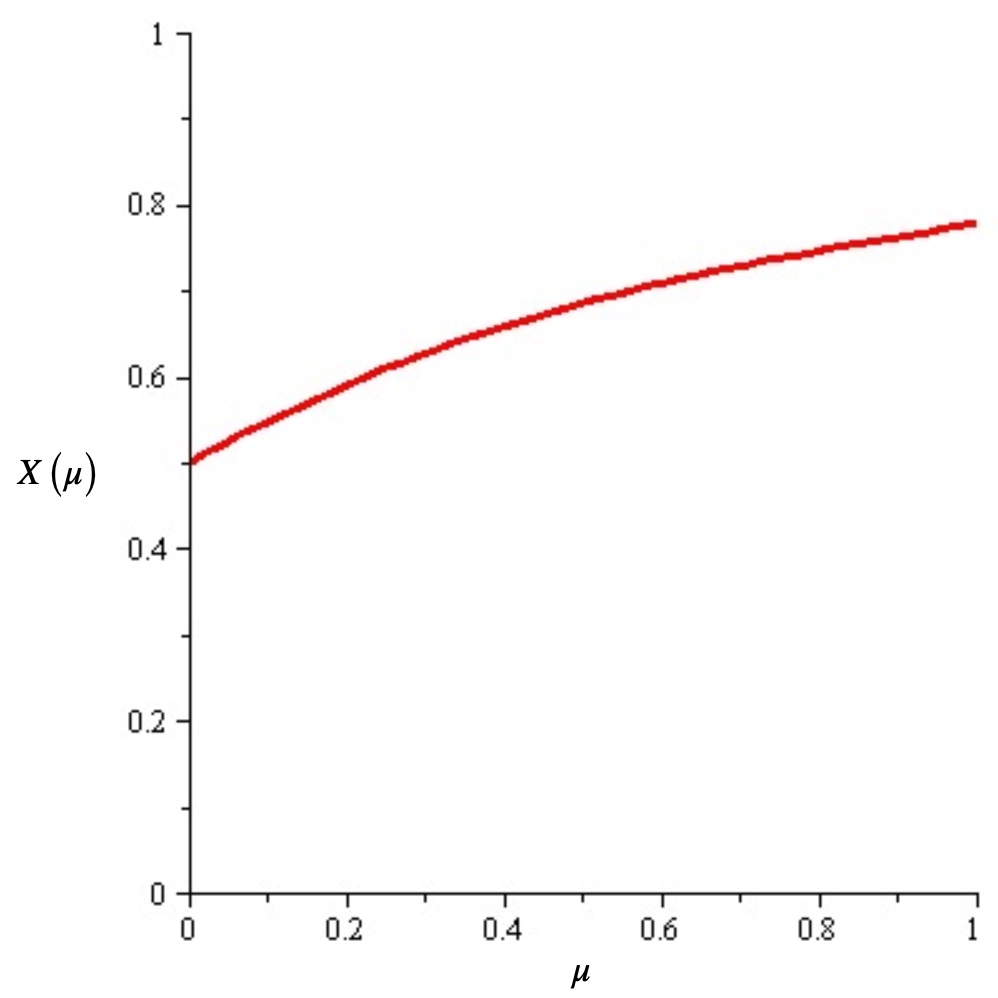} %%%
	\caption{Numerical solution for the autonomous system obtained from the mathematical software Maple.}
	\label{sol}
\end{figure}

Now in order to conduct a dynamical analysis, we investigate the stationary points of the differential equation \eqref{autoODE}. We first create a system of four differential equations by letting 
\begin{eqnarray}
    X' &=& Y, \nonumber\\
     Y' &=& Z, \nonumber\\
     Z' &=& W, \nonumber\\
    \label{fW}
    W' &=&f(X,Y,Z,W). 
\end{eqnarray}
Further using \eqref{autoODE}, $X''' \equiv W'$, the function $f$ in \eqref{fW} can be written explicitly giving
\begin{widetext}
\begin{eqnarray}
    W'  &=& \left\{-\frac{2}{3} \left(  -\frac{3}{2}X W   \left( - \left(X^2 Z   -2XYZ   +2 \right) ^{2} + \left( Y  
 \left( -4  Y  ^{2}+ \left( 4X +\frac{16}{3}\right) Y +  X ^{2}-\frac{11}{3}X  -\frac{4}{3}\right) -1\right) X  Z \right. \right.   \right. \nonumber\\
  &&    -4 \left( Y   \left( Y  -\frac{2}{3}
 \right)  \left( -\frac{1}{2} Y ^{2}+\frac{1}{6}\left( 6X  +3 \right) Y \right. -\frac{7}{6}X  -1\right)^{2} +\frac{3}{2} \left( 
-13XYZ +X^2 Z +9 \right)^{3}+6\frac { \left( XZ \right)^{4}}{Y  -1}  \nonumber\\
 &&+3 \left( Y \left( 8 Y ^{2}+ \left( X  -12 \right) Y  -\frac{3}{2}X
 +{\frac {35}{9}} \right) XZ -1 \right) ^{2} -3 \left( Y
   \left( \frac{1}{18} \left( 18 Y ^{2}-42Y  +24 \right) X  +6 Y^{3} \right. \right. \nonumber\\
&&\left. \left. \left. \left. \left. -{\frac {95}
{6}} Y ^{2}+\frac {233}{18}Y
  -\frac {29}{9} \right) -1 \right) ^{2}X  Z  +6 \left( Y  
 \left( Y  -\frac{2}{3}\right) \left( Y  -\frac{4}{3} \right)  \left( Y  -\frac{5}{6}\right)-1\right)^{4}\right)\right)\right\}   \Bigg/   \nonumber\\
 && \left\{  \left( Y   \left( X Z  -  Y ^{2}+\frac{5}{3}Y  -\frac{2}{3} \right) -1 \right)X^3\right\}.\label{wprime}
\end{eqnarray}
\end{widetext}
At the stationary points $X'=Y'=Z'=0$ which imply that $Y=Z=W=0$. Substituting $Y=Z=W=0$ into \eqref{wprime} and simplifying yields
\begin{align}
    W' &= \frac{40}{9X^3}.
\end{align}
Clearly $W'=0$ only if $X = (\mu+p) \to \infty$. This means that the stationary points for the autonomous system exist at infinity indicating the presence of a singularity there. 

%%%%%%%%%%%%%%%%%%%%%%%%%%%%%%%%%%%%%%%%%%%%%%%%%%%%%%%%%%%%%%%%%%%%%%%%%%%%%%%%%%%%%%%%%%%%%%%%%%%%%%%%%%%%%%%%%%%%%%
%%%%%%%%%%%%%%%%%%%%%%%%%%%%%%%%%%%%%%%%%%%%%%%%%%%%%%%%%%%%%%%%%%%%%%%%%%%%%%%%%%%%%%%%%%%%%%%%%%%%%%%%%%%%%%%%%%%%%%

\section{Discussion}
In this paper, we investigated the behaviour of the self-gravitating system of a shear-free perfect fluid with LRS-II symmetry through a geometrical perspective. We have provided a comprehensive answer to the question posed in the title of our paper, in the most generic covariant way, to extract the highly nonlinear differential equation that governs the equation of state. The novel results that emerged from our analysis can be summarized as follows:
\begin{itemize}
\item We can completely characterize all spherically symmetric shear-free perfect fluid spacetimes under a single classification that depends on the covariantly defined fluid acceleration and expansion.
\item One of the most important classes of shear-free spherical symmetry is the inhomogeneous dynamical class. We  pinpoint the exact constraint, given by a differential equation \eqref{autoODE}, governing the equation of state. It is interesting to note that the autonomous ODE \eqref{autoODE} obtained places a restriction solely on the equation of state. It does not place a restriction on any of the kinematical variables.  
\item Since the governing ODE is highly nonlinear in nature, we easily showed that usual equations of state used in astrophysical settings (for example, a linear equation of state or a finite combination of power law equations of state) will not satisfy this equation. Hence all the solutions belonging to this class are extremely special in terms of the matter content. We also presented a numerical solution for this ODE.
\end{itemize}

%%%%%%%%%%%%%%%%%%%%%%%%%%%%%%%%%%%%%%%%%%%%%%%%%%%%%%%%%%%%%%%%%%%%%%%%%%%%%%%%%%%%%%%%%%%%%%%%%%%%%%%%%%%%%%%%%%%%%%
%%%%%%%%%%%%%%%%%%%%%%%%%%%%%%%%%%%%%%%%%%%%%%%%%%%%%%%%%%%%%%%%%%%%%%%%%%%%%%%%%%%%%%%%%%%%%%%%%%%%%%%%%%%%%%%%%%%%%%

\begin{acknowledgements} 
CH is supported by the Oppenheimer Memorial Trust (OMT) and the University of KwaZulu-Natal. JH, RG and SDM are supported by the National Research Foundation (NRF), South Africa, and the University of KwaZulu-Natal. 
\end{acknowledgements}

%%%%%%%%%%%%%%%%%%%%%%%%%%%%%%%%%%%%%%%%%%%%%%%%%%%%%%%%%%%%%%%%%%%%%%%%%%%%%%%%%%%%%%%%%%%%%%%%%%%%%%%%%%%%%%%%%%%%%%
%%%%%%%%%%%%%%%%%%%%%%%%%%%%%%%%%%%%%%%%%%%%%%%%%%%%%%%%%%%%%%%%%%%%%%%%%%%%%%%%%%%%%%%%%%%%%%%%%%%%%%%%%%%%%%%%%%%%%%

\end{document}